\documentclass[twocolumn]{aastex631}

\usepackage{comment}
\usepackage[T1]{fontenc}

\newcommand\R[1]{#1}

\defcitealias{Allers2013}{AL13}

\shorttitle{Gravity Indices in Planet-hosting Ultracool Dwarfs}
\shortauthors{Davoudi et al.}

\begin{document}

\title{Gravity-sensitive Spectral Indices in Ultracool Dwarfs: Investigating Correlations with Metallicity and Planet Occurrence using SpeX and FIRE Observations}

\correspondingauthor{Fatemeh Davoudi}
\email{Fatemeh.Davoudi@uliege.be}

\author[0000-0002-1787-3444]{Fatemeh Davoudi}
\affiliation{Astrobiology Research Unit, Universit\'e de Li\`ege, All\'ee du 6 Ao\^ut 19C, B-4000 Li\`ege, Belgium}

\author[0000-0002-3627-1676]{Benjamin V.\ Rackham}
\affiliation{Department of Earth, Atmospheric and Planetary Science, Massachusetts Institute of Technology, 77 Massachusetts Avenue, Cambridge, MA 02139, USA}
\affiliation{Kavli Institute for Astrophysics and Space Research, Massachusetts Institute of Technology, Cambridge, MA, USA}

\author[0000-0003-2415-2191]{Julien de Wit}
\affiliation{Department of Earth, Atmospheric and Planetary Science, Massachusetts Institute of Technology, 77 Massachusetts Avenue, Cambridge, MA 02139, USA}

\author[0009-0004-8856-6793]{Jan Toomlaid}
\affiliation{Department of Earth, Atmospheric and Planetary Science, Massachusetts Institute of Technology, 77 Massachusetts Avenue, Cambridge, MA 02139, USA}

\author[0000-0003-1462-7739]{Michaël Gillon}
\affiliation{Astrobiology Research Unit, Universit\'e de Li\`ege, All\'ee du 6 Ao\^ut 19C, B-4000 Li\`ege, Belgium}

\author[0000-0002-5510-8751]{Amaury H.\ M.\ J.\ Triaud}
\affiliation{School of Physics \& Astronomy, University of Birmingham, Edgbaston, Birmingham B15 2TT, UK}

\author[0000-0002-6523-9536]{Adam J.\ Burgasser}
\affiliation{Department of Astronomy \& Astrophysics, University of California San Diego, La Jolla, CA 92093, USA}

\author[0000-0002-9807-5435]{Christopher A.\ Theissen}
\affiliation{Department of Astronomy \& Astrophysics, University of California San Diego, La Jolla, CA 92093, USA}



\begin{abstract}

We present a near-infrared spectroscopic analysis (0.9--2.4\,$\micron$) of gravity indices for 56 ultracool dwarfs (M5.5--L0), including exoplanet hosts SPECULOOS-2, SPECULOOS-3 and LHS\,3154. \R{Our dataset includes 59 spectra from the SpeX and FIRE spectrographs. We also discuss literature results for TRAPPIST-1.}
Using gravity-sensitive spectral indices, including FeH absorption (0.99, 1.20, and 1.55~$\micron$), the VO band at 1.06~$\micron$, the H-band continuum, and alkali lines such as K~I (1.17 and 1.25~$\micron$), we investigate correlations between \R{gravity classification}, stellar metallicity, and the presence of close-in transiting planets.
All four planet-hosting stars exhibit intermediate-gravity spectral signatures, despite indicators of field age.
However, a volume-corrected logistic regression reveals no significant association between gravity class and planet occurrence.
Among individual indices, we find FeH$_z$ to be the most promising tracer of planet-hosting status.
We tentatively identify a correlation between FeH$_z$ ($0.99\,\micron$) and planet presence at the $2\sigma$ level, though the result may reflect observational biases, including transit probability, small-number statistics, and detection sensitivity. 
More robustly, we find a significant anti-correlation between FeH$_z$ and [Fe/H] ($3.3\sigma$).
A Kruskal--Wallis test shows no significant [Fe/H] difference across gravity classes, suggesting the observed FeH$_z$--[Fe/H] trend is not driven by bulk metallicity differences.
We propose this anti-correlation may reflect the interplay between age, gravity, and composition: higher-metallicity objects may be systematically younger and have lower gravities, suppressing FeH absorption. 
While our results only hint at a link between gravity-related characteristics and planet occurrence among late-M dwarfs, they underscore the need for caution when using spectral diagnostics to infer the properties of planet-hosting ultracool dwarfs.

\end{abstract}



\section{Introduction} \label{sec:intro}

M dwarfs represent the most numerous stellar population in the galaxy, comprising approximately 75\% of all stars \citep{Henry2006, Henry2018}. Their low luminosities-ranging from 100 to 10,000 times less than the Sun—and masses spanning from about 0.61 $M_{\odot}$ for the hottest types down to the hydrogen-burning limit of 0.075 $M_{\odot}$ for the coolest \citep{Henry2024}, make them essential for studies of stellar formation, evolution, and galactic structure \citep{Reid2005}. Additionally, M dwarfs are prime targets for exoplanet searches due to their abundance and the favorable conditions they provide for detecting Earth-sized planets within their habitable zones \citep{Henry2024}.

Surface gravity significantly influences the atmospheric pressure and temperature profiles of M dwarfs, shaping the strength and morphology of molecular and atomic spectral features. In late-M dwarfs, gravity-sensitive indicators within the 0.9–2.5\,$\micron$ spectral range---such as FeH absorption bands (0.99, 1.20, and 1.55\,$\micron$), alkali lines including K\,\textsc{i} (1.17 and 1.25\,$\micron$) and Na\,\textsc{i} (1.14 and 2.21\,$\micron$), the VO band at 1.06\,$\micron$, and the shape of the H-band continuum---serve as diagnostic tools for youth and low gravity. Lower surface gravity reduces gas pressure in the photospheric layers, leading to weaker FeH and alkali line absorption, enhanced VO band strength, and a characteristic triangular H-band continuum shape \citep{Kleinmann1986, Joyce1998, Meyer1998, Lucas2001, Gorlova2003, McGovern2004, Allers2007, Lodieu2008, Allers2013}. 
Expanding on earlier efforts that introduced empirical gravity classifications based on optical alkali and CaH and VO features \citep{Kirkpatrick2008, Cruz2009}, \cite{Allers2013} (hereafter \citetalias{Allers2013}) systematically quantified these near-infrared (NIR) features across spectral types M5--L7, enabling the development of gravity indices that quantitatively assess youth in ultracool dwarfs.

While gravity-sensitive spectral indices are well established for young brown dwarfs and field M dwarfs, their interpretation in exoplanet-hosting M dwarfs remains less explored. Intriguingly, some field-age M dwarfs with known planetary systems, such as TRAPPIST-1 and Teegarden’s Star, exhibit spectral signatures commonly associated with youth and low gravity, despite their older estimated ages.

TRAPPIST-1, an M8 dwarf hosting seven terrestrial planets with orbital periods ranging from 1.5 to 18.8 days \citep{Gillon2017}, displays weak FeH absorption and a triangular $H$-band continuum—traits typically attributed to low gravity \citep[e.g.,][]{Gillon2016, Burgasser2017}. However, TRAPPIST-1 is classified as a field-age star \citep[e.g.,][]{Gizis2002, Filippazzo2015, Gonzales2019} with an estimated age of $7.6 \pm 2.2$\,Gyr \citep{Burgasser2017}. Its overall spectral energy distribution also resembles older field stars \citep{Gonzales2019}. Possible explanations for its youth-like spectral features include magnetic activity or tidal interactions with its planets \citep{Gonzales2019}, potentially leading to radius inflation via inhibited convection \citep{MacDonald2017}.

A similar ambiguity is seen in Teegarden’s Star, a magnetically inactive M7.5 dwarf at 3.85 pc \citep{Henry2006, Teegarden2003}. Hosting at least three exoplanets with orbital periods of 4.906, 11.416, and 26.13 days, including two within the conservative habitable zone, it is estimated to be older than 8 Gyr \citep{Zechmeister2019, Dreizler2024}. Nevertheless, it has been classified as an M7.5 $\beta$ dwarf, indicating intermediate gravity and youth-like spectral characteristics \citep{Gagne2015}.

These cases raise the possibility that the unusual gravity signatures seen in these ultracool dwarfs may be connect to their planet-hosting status. 
One explanation is that higher stellar metallicity inflates stellar radii by increasing atmospheric opacity, thereby mimicking low-gravity features and potentially influencing both the star's spectral appearance and its propensity to form planets. 
However, metallicity can also directly affect gravity-sensitive features:
metal-poor subdwarfs have been shown to exhibit triangular $H$-band continua and altered FeH absorption that resemble signatures of youth \citep{Aganze2016, Martin2017}.
Similar spectral features have also been reported in ${\sim}10\%$ of nearby field-age M dwarfs without known planets \citep{BardalezGagliuffi2019}, suggesting an impact from other stellar properties (beyond age or planet occurrence)---or indicating that many nearby M-dwarfs host undetected planets.

While theoretical models predict a decline in super-Earth occurrence toward the lowest-mass stars \citep{Mulders2021}, which agrees with occurrence rates from \textit{K2} \citep{Sestovic2020}, a global analysis of the TRAPPIST Ultracool Dwarf Transit Survey finds a planet occurrence rate of $\gtrsim10\%$ for short-period, Earth-sized planets around ultracool dwarfs \citep{Lienhard2020}.
Given our generally poor constraints on planet occurrence around late-M dwarfs and the intriguing pattern of low-gravity spectroscopic indicators for detected systems, there exists a notable lack of systematic observational studies directly comparing gravity-sensitive features, metallicity, and planet occurrence within a sample of late-M dwarfs.

To investigate these trends, here we analyze a sample of 56 ultracool M dwarfs spanning spectral types M5.5 to L0, including the known exoplanet hosts \R{ SPECULOOS-2, SPECULOOS-3, and LHS~3154. For comparison, we also consider TRAPPIST-1.} Using ground-based NIR spectroscopy, we examine gravity-sensitive features to classify each object by gravity regime (high, intermediate, or low) and explore potential correlations with metallicity and planet occurrence. Our aim is to determine whether unusual gravity signatures in planet-hosting late-M dwarfs are linked to underlying physical properties relevant to planet formation and detectability.

This paper is structured as follows. \autoref{observation} describes our sample selection, observations, and data reduction. \autoref{Methods} details the computation and analysis of gravity-sensitive indices. \autoref{results} presents our results and discusses gravity classification in exoplanet-hosting stars. Finally, \autoref{conclusion} summarizes our key findings and implications.

\section{Sample} \label{observation}

We analyze 59 spectra from 56 late-M ultracool dwarfs of spectral types M5.5 to L0 (\autoref{table1}).
These targets were selected as part of an ongoing campaign to spectroscopically classify candidates for the SPECULOOS transit survey \citep{Sebastian2021}.
Since our study focuses on gravity-sensitive spectral indices, we limited the sample to field-age stars to avoid confusion with youth-induced low-gravity features.
We used the BANYAN $\Sigma$ tool\footnote{\url{https://www.exoplanetes.umontreal.ca/banyan/banyansigma.php}} \citep{Gagne2018} to assess kinematic membership in 27 known young associations within 150\,pc, excluding all likely members.

\subsection{Observations and Data Reduction}

Near-infrared spectroscopic observations of our sample were conducted using the SpeX spectrograph \citep{Rayner2003} on the 3.2-meter NASA Infrared Telescope Facility (IRTF) at Mauna Kea, Hawaii, and the Folded-port InfraRed Echellette (FIRE) spectrograph \citep{Simcoe2013} on the 6.5-meter Magellan Baade Telescope at Las Campanas Observatory, Chile.\R{ Of the 59 spectra analyzed, 39 were obtained with SpeX and 20 with FIRE (\autoref{table1}). Additionally, we include one FIRE and one SpeX spectrum of TRAPPIST-1.} Except for spectra associated with the exoplanet-hosting stars TRAPPIST-1, SPECULOOS-2, and SPECULOOS-3, none have been previously published.

\begin{deluxetable*}{lllll}
\tablecaption{Summary of Spectroscopic Observations and Spectral Types \label{table1}}
\tablewidth{0pt}
\tabletypesize{\scriptsize}
\tablehead{
\colhead{Object} & \colhead{Instrument} & \colhead{Obs. Date} & \colhead{Spectral Type} & \colhead{SNR (per pixel)}
}
\startdata
2MASS J00202922+3305081 & SpeX & 2021/08/30 & M5.5V ($\pm$0.5) & 62.6 \ \\
2MASS J00251602+5422547 & SpeX & 2022/08/12 & M6.5V ($\pm$0.5) & 72.3 \ \\
2MASS J02195603+5919273 & SpeX & 2022/08/12 & M5.5V ($\pm$0.5) & 96.7 \ \\
2MASS J02224767-2732349 & FIRE & 2023/02/01 & M6.5V ($\pm$0.5) & 95.72 \ \\
2MASS J03544620+2416246 & SpeX & 2021/08/30 & M6V ($\pm$1.0) & 52.8 \ \\
2MASS J04164276+1310587 & SpeX & 2021/12/24 & M6V ($\pm$1.0)) & 86.4 \ \\
2MASS J04333002+5635320 & SpeX & 2022/02/11 & M8V ($\pm$0.5) & 67.1 \ \\
2MASS J04393407-3235516 & FIRE & 2021/01/07 & M6V ($\pm$0.5) & 165.0 \ \\
2MASS J04490464+5138412 & SpeX & 2022/02/08 & M6.5V ($\pm$0.5) & 76.9 \ \\
2MASS J04511406+0305285 & FIRE & 2021/01/08 & M7V ($\pm$1.0) & 126.3 \\
2MASS J04511406+0305285 & SpeX & 2021/12/24 & M8V ($\pm$0.5) & 43.9 \ \\
2MASS J04513734-5818519 & FIRE & 2021/01/07 & M6V ($\pm$0.5) & 150.0 \ \\
2MASS J05220976+5754046 & SpeX & 2022/02/08 & M8.5V ($\pm$0.5) & 56.6 \ \\
2MASS J05335379+5054170 & SpeX & 2022/02/11 & M8V ($\pm$0.5) & 58.8 \ \\
2MASS J05512511+5511208 & SpeX & 2022/02/08 & M6V ($\pm$1.0) & 77.7 \ \\
2MASS J06020172-1001565 & FIRE & 2021/01/08 & M6.5V ($\pm$0.5) & 141.3 \ \\
2MASS J06431389+1631428 & FIRE & 2021/01/08 & M7V ($\pm$0.5) & 159.6 \\
2MASS J06431389+1631428 & SpeX & 2021/12/24 & M7V ($\pm$0.5) & 49.5 \ \\
2MASS J07552745-2404374 & FIRE & 2021/01/07 & M6.5V ($\pm$0.5) & 130.8 \ \\
2MASS J08055713+0417035 & FIRE & 2021/01/08 & M6.5V ($\pm$0.5) & 197.7 \\
2MASS J08055713+0417035 & SpeX & 2021/12/24 & M6V ($\pm$1.0) & 90.9 \ \\
2MASS J08330310+3706083 & SpeX & 2021/12/24 & M8V ($\pm$0.5) & 71.1 \ \\
2MASS J08334323-5336417 & FIRE & 2021/01/07 & M7V ($\pm$0.5) & 152.9 \ \\
2MASS J09332510-4353384 & FIRE & 2022/06/14 & M6V ($\pm$1.0) & 103.6 \ \\
2MASS J09332625-4353366 & FIRE & 2022/06/14 & M6V ($\pm$1.0) & 104.2\ \\
2MASS J09365564-2609422 & SpeX & 2022/02/08 & M8V ($\pm$0.5) & 61.6 \ \\
2MASS J09432994-3833560 & SpeX & 2022/02/11 & M6V ($\pm$1.0) & 79.4 \ \\
2MASS J10424135-2416050 & SpeX & 2021/12/24 & M6.5V ($\pm$0.5) & 25.9 \ \\
2MASS J10542786-5431322 & FIRE & 2022/06/14 & M6.5V ($\pm$1.0) & 85.1\ \\
2MASS J11155037-6731332 & FIRE & 2022/06/14 & M8V ($\pm$0.5) & 114.0\ \\
2MASS J11231964-0509045 & SpeX & 2022/02/08 & M6V ($\pm$0.5) & 69.4 \ \\
2MASS J12294530+0752379 & SpeX & 2022/02/11 & M6.5V ($\pm$0.5) & 81.4\ \\
2MASS J13273095+0149384 & SpeX & 2022/02/11 & M6.5V ($\pm$0.5) & 69.0 \ \\
2MASS J13313937-6513056 & FIRE & 2022/06/14 & L0 ($\pm$1.0) & 76.3 \ \\
2MASS J14230252+5146303 & SpeX & 2022/04/19 & M7V ($\pm$0.5) & 52.3 \ \\
2MASS J14253465+2540050 & SpeX & 2022/04/19 & M6V ($\pm$0.5) & 85.0 \ \\
2MASS J16105843-0631325 & SpeX & 2022/07/21 & M5.5V ($\pm$0.5) & 82.2 \ \\
2MASS J16210447-3711373 & SpeX & 2022/04/15 & M7V ($\pm$1.0) & 60.8 \ \\
2MASS J17120433-0323300 & SpeX & 2022/04/19 & M6V ($\pm$0.5) & 88.7 \ \\
2MASS J17364180-3425459 & FIRE & 2022/06/15 & M6V ($\pm$1.0) & 124.5 \ \\
2MASS J17415439+0940537 & SpeX & 2022/07/21 & M7V ($\pm$0.5) & 67.2 \ \\
2MASS J18365842-3507176 & FIRE & 2022/06/15 & M6.5V ($\pm$0.5) & 74.0 \ \\
2MASS J18485108-8214422 & FIRE & 2022/06/14 & M7V ($\pm$1.0) & 93.4 \ \\
2MASS J18545092-5704417 & FIRE & 2022/06/14 & M7V ($\pm$1.0) & 104.9 \ \\
2MASS J19212977-2915507 & SpeX & 2021/08/30 & M7V ($\pm$0.5) & 89.8 \ \\
2MASS J19332754+2150009 & SpeX & 2022/07/21 & M8.5V ($\pm$0.5) & 51.9 \ \\
2MASS J19395199-5750339 & FIRE & 2022/06/14 & M6.5V ($\pm$0.5) & 67.3 \ \\
2MASS J19544358+1801581 & SpeX & 2021/08/30 & M8V ($\pm$0.5) & 72.4 \ \\
2MASS J20125255+1246315 & SpeX & 2021/10/19 & M6.5V ($\pm$0.5) & 57.1 \ \\
2MASS J20291194+5750317 & SpeX & 2021/08/30 & M6V ($\pm$0.5) & 67.0 \ \\
2MASS J20495272-1716083 & SpeX & 2021/10/19 & M6.5V ($\pm$0.5) & 47.1 \ \\
2MASS J21010483+0307047 & SpeX & 2021/10/19 & M6.5V ($\pm$0.5) & 47.2 \ \\
2MASS J21265788+2531080 & SpeX & 2021/08/30 & M8.5V ($\pm$0.5) & 52.4 \ \\
2MASS J21381698+5257188 & SpeX & 2022/08/12 & M7V ($\pm$1.0) & 66.4 \ \\
2MASS J21513137-4017229 & FIRE & 2022/06/14 & M7.5V ($\pm$1.0) & 116.2 \ \\
2MASS J22244238+2230425 & SpeX & 2021/08/30 & M6V ($\pm$1.0) & 49.7 \ \\
SPECULOOS-2 & SpeX & 2021/08/30 & M6V($\pm$0.5) & 100.0 \ \\
SPECULOOS-3 & SpeX & 2021/08/30 & M6.5V($\pm$0.5) & 54.8 \ \\
LHS 3154 & SpeX & 2025/04/02 & M6V ($\pm$1.0) & 64.6 \ \\
\enddata
\end{deluxetable*}

\subsubsection{IRTF/SpeX Near-IR Spectroscopy}

SpeX observations were conducted in the short-wavelength cross-dispersed (SXD) mode, covering the 0.8--2.4\,$\micron$ range at a resolving power of $R {\approx} 2000$ using the 0.3$\arcsec$ slit. Calibration frames, including dome flat-field frames and ThAr arc lamp spectra, were used to correct pixel-to-pixel sensitivity variations and to perform wavelength calibration. The data were reduced with the \texttt{SPEXTOOL} v4.1 pipeline \citep{Cushing2004, Vacca2003}, which applies flat-field correction, sky subtraction, spectral extraction, telluric correction, and flux calibration. The sky background was removed by differencing ABBA nodding pairs along the slit, and one-dimensional spectra were extracted from the two-dimensional images. Telluric absorption features were corrected using the \texttt{xtellcor} routine \citep{Vacca2003} with A0\,V standard stars observed at similar airmasses. These standards were also used to perform the flux calibration.

\subsubsection{FIRE Near-IR Spectroscopy}

FIRE observations were conducted in the echellette mode, covering the full 0.8--2.5\,$\micron$ wavelength range at a resolving power of $R {\approx} 6000$. The setup included the 0.45$\arcsec$ slit, high-gain mode, and sample-up-the-ramp readout. ABBA nodding along the slit was used to enable sky subtraction. Quartz lamp flats and ThAr lamps were used for flat-field correction and wavelength calibration, respectively. Observations were conducted at airmasses below 1.7 under typical seeing conditions of 0.7$\arcsec$ to 1.1$\arcsec$. Data reduction and spectral extraction were performed using the \texttt{FIREHOSE} pipeline \citep{Simcoe2013}, which calibrates and combines the echelle orders into one-dimensional spectra. Telluric correction and flux calibration were performed using the \texttt{xtellcor} package \citep{Vacca2003} and observations of A0\,V standard stars observed at similar airmasses.

\subsubsection{A Note on Resolution}

To evaluate whether gravity indices are influenced by differences in spectral resolution between SpeX ($R{\approx}2000$) and FIRE ($R{\approx}6000$), we refer to the findings of \citetalias{Allers2013}. They analyzed low-resolution ($R{\approx}100$) and moderate-resolution ($R{\approx}750–2000$) spectra separately and demonstrated that, while one spectral index is less discernible in low-resolution data, their gravity scoring method remains consistent across resolutions. Similarly, \cite{Gonzales2019} analyzed both low- and medium-resolution spectra ($\lambda/\Delta\lambda > 1000$ at the $J$-band) and found gravity class designations to be consistent across resolutions, with only a single outlier. These findings support the reliability of using moderate-resolution spectra for gravity analysis. Given that our spectra fall within the moderate-resolution range ($R{\approx}2000–6000$), they are thus reliable for use in our comparative gravity studies.

\subsection{Spectral Classification}

\begin{figure*}[!htp]
	\centering
	\includegraphics[width=\textwidth, height=\textheight, keepaspectratio]{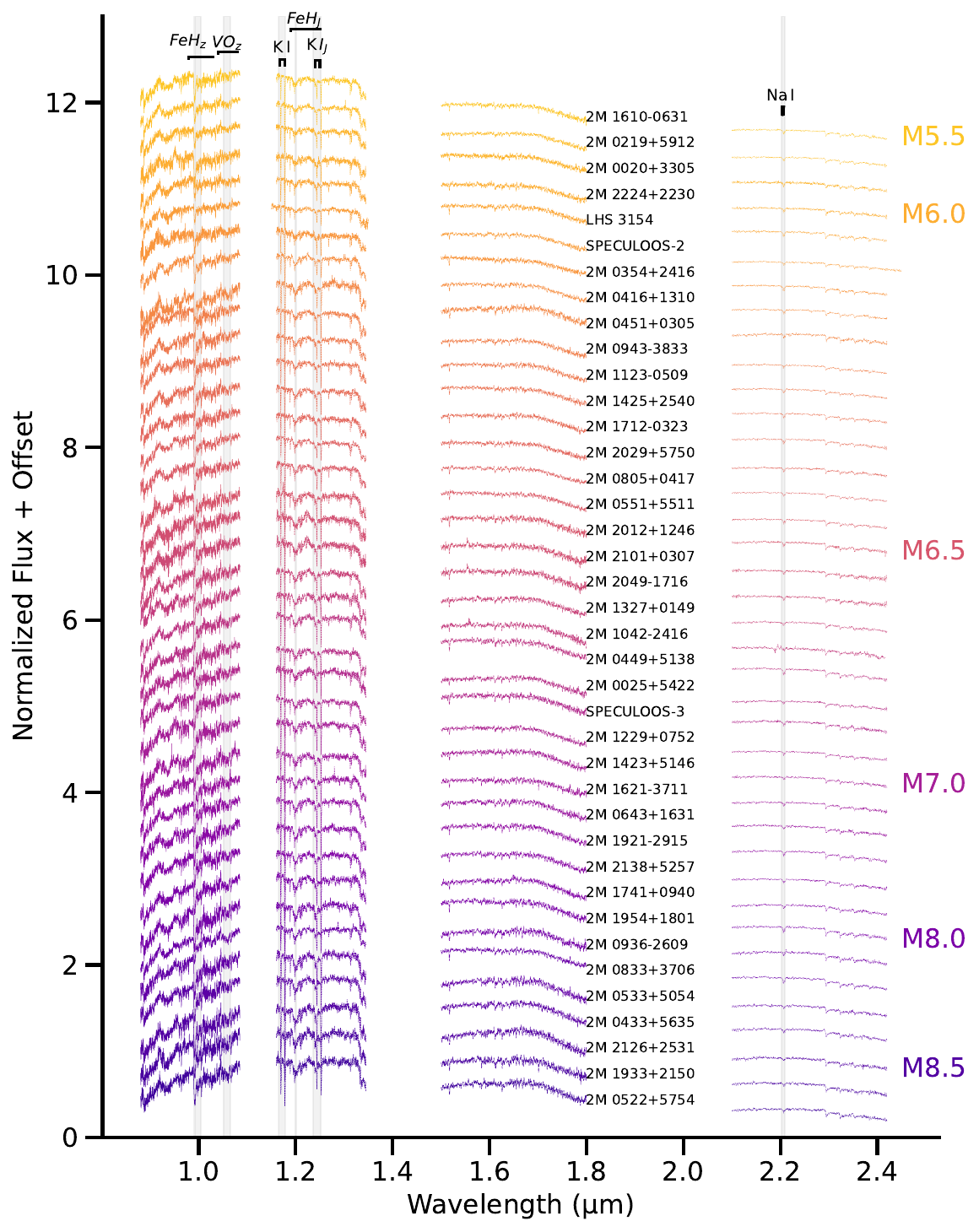}
	\caption{Near-IR SpeX/SXD spectra ($R {\sim} 2000$) of the M dwarfs in our sample. Spectral types and positions of significant gravity-sensitive features are indicated. \R{These spectra are publicly available on Zenodo: \href{https://doi.org/10.5281/zenodo.16420633}{10.5281/zenodo.16420633}.}} 
	\label{figure1}
\end{figure*}

\begin{figure*}[!htp]
	\centering
	\includegraphics[width=\textwidth, height=\textheight, keepaspectratio]{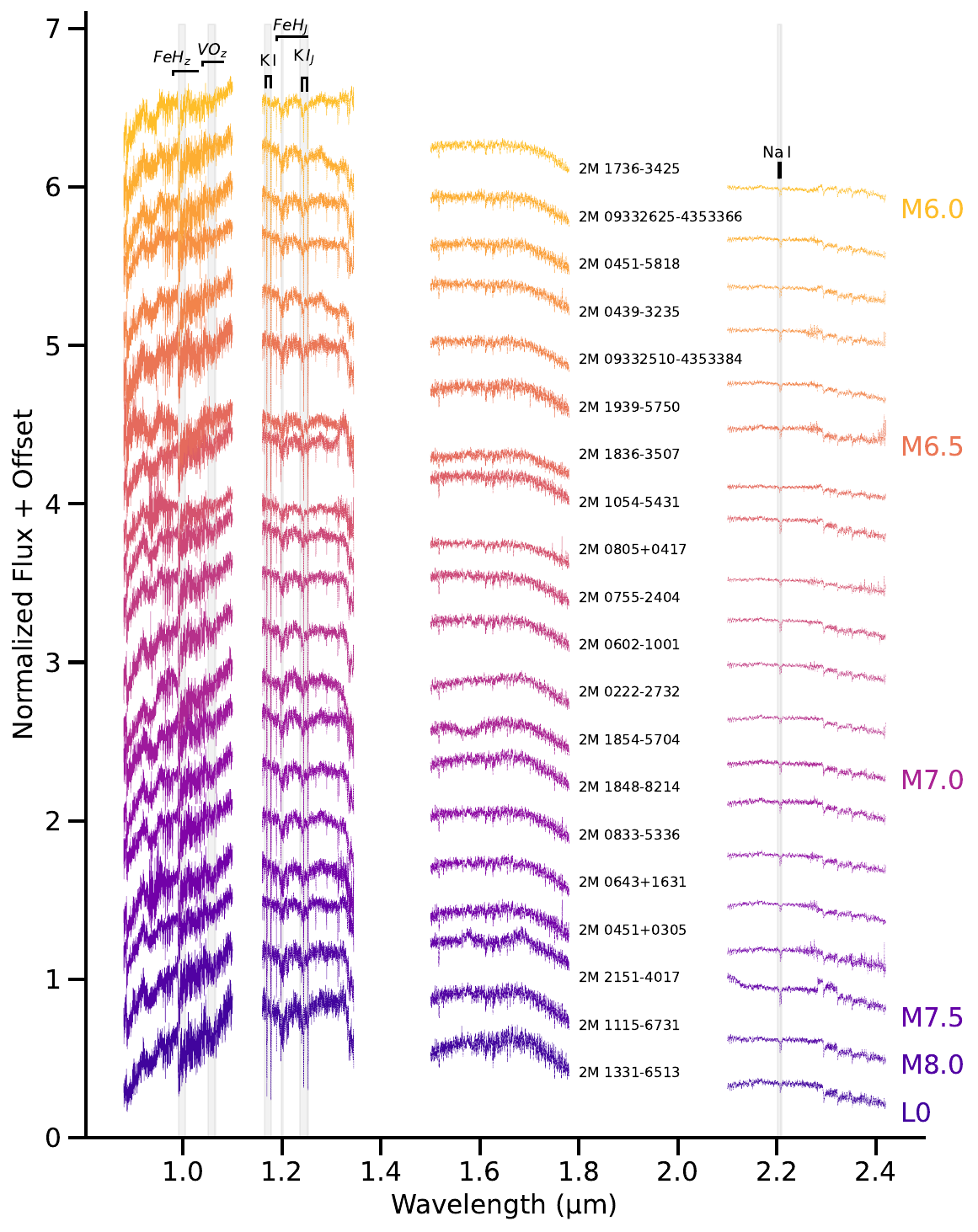}
	\caption{Near-IR FIRE spectra ($R {\sim} 6000$) of the M dwarfs in our sample. Spectral types and positions of significant gravity-sensitive features are indicated. \R{These spectra are publicly available on Zenodo: \href{https://doi.org/10.5281/zenodo.16420633}{10.5281/zenodo.16420633}.}} 
	\label{figure2}
\end{figure*}

SpeX and FIRE spectra of our sample are shown in \autoref{figure1} and \autoref{figure2}, respectively. Details for each target—including the instrument used, observation date, near-infrared spectral type, and median signal-to-noise ratio (SNR)—are provided in \autoref{table1}. The median per-pixel SNR was calculated over the 8000–24000~\AA\ wavelength range, excluding telluric-dominated regions between 10850–11250~\AA\ and 19000–21000~\AA. All spectra in our sample have a median per-pixel $\mathrm{SNR} > 50$. Each reduced spectrum was visually inspected to confirm the presence and consistency of key ultracool dwarf features.

We assigned near-infrared spectral types by comparing each spectrum to those in the IRTF Spectral Library \citep{Cushing2005, Rayner2009}, focusing on the 0.9--1.4\,$\micron$ range, following \citet{Kirkpatrick2010}. 
Both target and template spectra were normalized to their median flux, and reduced $\chi^2$ values were computed to identify the best-matching templates.
The assigned spectral type corresponds to the top match, while the uncertainty ($\pm0.5$ or $\pm1.0$ subtype) reflects the separation between the best and second-best fits.
Classifications were validated through visual inspection.
We also testing an alternative method that removed linear slopes between spectra before comparison, which had no notable effect.

\begin{deluxetable*}{llllllll}
\tablecaption{Gravity Indices and Classifications \label{table2}}
\tablewidth{0pt}
\tabletypesize{\scriptsize}
\tablehead{
\colhead{Object (2MASS J)} & \colhead{FeH\(_z\) index} & \colhead{FeH\(_J\) index} & \colhead{VO\(_z\) index} & \colhead{KI\(_J\) index} & \colhead{H-cont index} & \colhead{Gravity Score} & \colhead{Gravity Class}
}
\startdata
00202922+3305081 & 1.069 ± 0.014 & 1.050 ± 0.010 & 1.016 ± 0.012 & 1.046 ± 0.011 & 0.988 ± 0.009 & 0n01 & FLD-G	\ \\
00251602+5422547 & 1.149 ± 0.015 & 1.101 ± 0.009 & 0.998 ± 0.011 & 1.086 ± 0.011 & 0.983 ± 0.006 & 0n01 & FLD-G	\ \\
02195603+5919273 & 1.074 ± 0.011 & 1.050 ± 0.007 & 1.036 ± 0.009 & 1.053 ± 0.009 & 0.993 ± 0.008 & 0n01 & FLD-G	\ \\
02224767-2732349 & 1.157 ± 0.017 & 1.107 ± 0.009 & 1.053 ± 0.010 & 1.084 ± 0.011 & 0.980 ± 0.007 & 0n00 & FLD-G	\ \\
03544620+2416246 & 1.071 ± 0.026 & 1.058 ± 0.014 & 1.022 ± 0.022 & 1.055 ± 0.014 & 0.992 ± 0.008 & 0n01 & FLD-G	\ \\
04164276+1310587 & 1.132 ± 0.013 & 1.098 ± 0.008 & 1.013 ± 0.010 & 1.058 ± 0.010 & 0.984 ± 0.005 & 0n00 & FLD-G	\ \\
04333002+5635320 & 1.214 ± 0.016 & 1.132 ± 0.010 & 1.019 ± 0.012 & 1.081 ± 0.011 & 0.980 ± 0.007 & 0n01 & FLD-G	\ \\
04393407-3235516 & 1.094 ± 0.010 & 1.071 ± 0.005 & 1.010 ± 0.007 & 1.057 ± 0.008 & 0.982 ± 0.005 & 0n00 & FLD-G	\ \\
04490464+5138412 & 1.133 ± 0.014 & 1.112 ± 0.008 & 1.033 ± 0.011 & 1.072 ± 0.010 & 0.977 ± 0.009 & 0n00 & FLD-G	\ \\
04511406+0305285(FIRE) & 1.169 ± 0.026 & 1.103 ± 0.014 & 1.047 ± 0.019 & 1.083 ± 0.016 & 0.981 ± 0.007 & 0n01 & FLD-G	\ \\
04511406+0305285(SpeX) & 1.146 ± 0.013 & 1.121 ± 0.007 & 1.062 ± 0.009 & 1.081 ± 0.009 & 0.983 ± 0.006 & 1n01 & INT-G	\ \\
04513734-5818519 & 1.168 ± 0.011 & 1.096 ± 0.006 & 1.021 ± 0.007 & 1.077 ± 0.008 & 0.976 ± 0.004 & 0n00 & FLD-G	\ \\
05220976+5754046 & 1.243 ± 0.025 & 1.157 ± 0.012 & 1.052 ± 0.017 & 1.077 ± 0.014 & 0.975 ± 0.005 & 1n11 & INT-G	\ \\
05335379+5054170 & 1.281 ± 0.019 & 1.161 ± 0.011 & 1.009 ± 0.013 & 1.092 ± 0.012 & 0.961 ± 0.009 & 0n00 & FLD-G	\ \\
05512511+5511208 & 1.121 ± 0.013 & 1.094 ± 0.008 & 1.013 ± 0.011 & 1.064 ± 0.010 & 0.976 ± 0.007 & 0n00 & FLD-G	\ \\
06020172-1001565 & 1.127 ± 0.011 & 1.085 ± 0.006 & 1.007 ± 0.007 & 1.053 ± 0.008 & 0.982 ± 0.006 & 0n11 & INT-G	\ \\
06431389+1631428(FIRE) & 1.212 ± 0.022 & 1.119 ± 0.012 & 1.009 ± 0.016 & 1.088 ± 0.014 & 0.980 ± 0.007 & 0n01 & FLD-G	\ \\
06431389+1631428(SpeX) & 1.168 ± 0.011 & 1.111 ± 0.006 & 1.006 ± 0.007 & 1.070 ± 0.008 & 0.984 ± 0.005 & 0n00 & FLD-G	\ \\
07552745-2404374 & 1.091 ± 0.012 & 1.077 ± 0.006 & 1.036 ± 0.008 & 1.056 ± 0.009 & 0.980 ± 0.006 & 1n10 & INT-G	\ \\
08055713+0417035(FIRE) & 1.139 ± 0.012 & 1.080 ± 0.007 & 0.994 ± 0.009 & 1.055 ± 0.009 & 0.979 ± 0.008 & 0n11 & INT-G	\ \\
08055713+0417035(SpeX) & 1.130 ± 0.009 & 1.085 ± 0.005 & 1.009 ± 0.007 & 1.053 ± 0.007 & 0.981 ± 0.005 & 0n00 & FLD-G	\ \\
08330310+3706083 & 1.139 ± 0.015 & 1.108 ± 0.009 & 1.062 ± 0.012 & 1.072 ± 0.011 & 0.971 ± 0.009 & 1n11 & INT-G	\ \\
08334323-5336417 & 1.121 ± 0.011 & 1.093 ± 0.006 & 1.057 ± 0.008 & 1.069 ± 0.008 & 0.975 ± 0.005 & 0n00 & FLD-G	\ \\
09332510-4353384 & 1.138 ± 0.016 & 1.110 ± 0.008 & 0.998 ± 0.009 & 1.070 ± 0.011 & 0.972 ± 0.007 & 0n00 & FLD-G	\ \\
09332625-4353366 & 1.164 ± 0.015 & 1.114 ± 0.008 & 1.003 ± 0.009 & 1.062 ± 0.010 & 0.972 ± 0.007 & 0n00 & FLD-G	\ \\
09365564-2609422 & 1.282 ± 0.021 & 1.170 ± 0.011 & 1.029 ± 0.014 & 1.102 ± 0.012 & 0.967 ± 0.005 & 0n01 & FLD-G	\ \\
09432994-3833560 & 1.152 ± 0.013 & 1.083 ± 0.008 & 1.001 ± 0.010 & 1.064 ± 0.010 & 0.983 ± 0.006 & 0n00 & FLD-G	\ \\
10424135-2416050 & 1.150 ± 0.025 & 1.112 ± 0.020 & 1.036 ± 0.023 & 1.069 ± 0.022 & 0.970± 0.012 & 0n00 & FLD-G	\ \\
10542786-5431322 & 1.103 ± 0.017 & 1.086 ± 0.009 & 1.040 ± 0.011 & 1.068 ± 0.012 & 0.983 ± 0.008 & 1n01 & INT-G	\ \\
11155037-6731332 & 1.142 ± 0.015 & 1.115 ± 0.008 & 1.061 ± 0.010 & 1.085 ± 0.010 & 0.985 ± 0.006 & 1n02 & INT-G	\ \\
11231964-0509045 & 1.148 ± 0.015 & 1.086 ± 0.009 & 1.002 ± 0.011 & 1.063 ± 0.011 & 0.978 ± 0.010 & 0n00 & FLD-G	\ \\
12294530+0752379 & 1.154 ± 0.013 & 1.088 ± 0.008 & 0.998 ± 0.010 & 1.058 ± 0.010 & 0.979 ± 0.009 & 0n10 & FLD-G	\ \\
13273095+0149384 & 1.176 ± 0.016 & 1.107 ± 0.009 & 1.012 ± 0.012 & 1.071 ± 0.011 & 0.993 ± 0.010 & 0n02 & INT-G	\ \\
13313937-6513056 & 1.241 ± 0.024 & 1.198 ± 0.011 & 1.119 ± 0.008 & 1.119 ± 0.010 & 0.969 ± 0.008 & 1001 & INT-G	\ \\
14230252+5146303 & 1.204 ± 0.017 & 1.081 ± 0.011 & 1.009 ± 0.013 & 1.068 ± 0.013 & 0.996 ± 0.009 & 0n02 & INT-G	\ \\
14253465+2540050 & 1.122 ± 0.012 & 1.082 ± 0.008 & 1.013 ± 0.010 & 1.055 ± 0.010 & 0.981 ± 0.007 & 0n00 & FLD-G	\ \\
16105843-0631325 & 1.112 ± 0.013 & 1.078 ± 0.008 & 1.005 ± 0.010 & 1.061 ± 0.010 & 0.976 ± 0.009 & 0n00 & FLD-G	\ \\
16210447-3711373 & 1.191 ± 0.019 & 1.129 ± 0.011 & 1.032 ± 0.014 & 1.089 ± 0.012 & 0.979 ± 0.005 & 0n00 & FLD-G	\ \\
17120433-0323300 & 1.126 ± 0.012 & 1.092 ± 0.008 & 1.019 ± 0.010 & 1.068 ± 0.010 & 0.977± 0.008 & 0n00 & FLD-G	\ \\
17364180-3425459 & 1.082 ± 0.013 & 1.084 ± 0.007 & 1.019 ± 0.009 & 1.068 ± 0.009 & 0.982 ± 0.006 & 0n00 & FLD-G	\ \\
17415439+0940537 & 1.184 ± 0.015 & 1.111 ± 0.009 & 1.009 ± 0.012 & 1.075 ± 0.011 & 0.979 ± 0.006 & 0n00 & FLD-G	\ \\
18365842-3507176 & 1.279 ± 0.027 & 1.123 ± 0.011 & 0.977 ± 0.013 & 1.108 ± 0.013 & 0.966 ± 0.007 & 0n00 & FLD-G	\ \\
18485108-8214422 & 1.118 ± 0.016 & 1.097 ± 0.008 & 1.028 ± 0.010 & 1.082 ± 0.011 & 0.978 ± 0.007 & 0n00 & FLD-G	\ \\
18545092-5704417 & 1.281 ± 0.018 & 1.101 ± 0.008 & 1.018 ± 0.009 & 1.084 ± 0.010 & 0.984 ± 0.007 & 0n01 & FLD-G	\ \\
19212977-2915507 & 1.170 ± 0.012 & 1.118 ± 0.008 & 1.023 ± 0.010 & 1.072 ± 0.009 & 0.992 ± 0.008 & 0n02 & INT-G	\ \\
19332754+2150009 & 1.254 ± 0.026 & 1.159 ± 0.012 & 1.075 ± 0.018 & 1.104 ± 0.014 & 0.962 ± 0.012 & 1n01 & INT-G	\ \\
19395199-5750339 & 1.103 ± 0.025 & 1.099 ± 0.012 & 1.038 ± 0.014 & 1.076 ± 0.014 & 0.986 ± 0.009 & 1n01 & INT-G	\ \\
19544358+1801581 & 1.176 ± 0.016 & 1.136 ± 0.009 & 1.058 ± 0.012 & 1.078 ± 0.011 & 0.986 ± 0.009 & 0n02 & INT-G	\ \\
20125255+1246315 & 1.180 ± 0.019 & 1.101 ± 0.011 & 1.024 ± 0.015 & 1.075 ± 0.013 & 0.993 ± 0.012 & 0n02 & INT-G	\ \\
20291194+5750317 & 1.123 ± 0.015 & 1.080 ± 0.010 & 0.995 ± 0.012 & 1.061 ± 0.011 & 0.979 ± 0.009 & 0n00 & FLD-G	\ \\
20495272-1716083 & 1.147 ± 0.022 & 1.096 ± 0.013 & 1.001 ± 0.017 & 1.074 ± 0.015 & 0.978 ± 0.007 & 0n00 & FLD-G	\ \\
21010483+0307047 & 1.140 ± 0.022 & 1.097 ± 0.013 & 1.003 ± 0.017 & 1.075 ± 0.015 & 0.980 ± 0.014 & 0n00 & FLD-G	\ \\
21265788+2531080 & 1.277 ± 0.024 & 1.163 ± 0.013 & 1.039 ± 0.017 & 1.104 ± 0.014 & 0.972 ± 0.010 & 0n01 & FLD-G	\ \\
21381698+5257188 & 1.198 ± 0.016 & 1.139 ± 0.010 & 1.008 ± 0.012 & 1.076 ± 0.011 & 0.987 ± 0.006 & 0n01 & FLD-G	\ \\
21513137-4017229 & 1.097 ± 0.013 & 1.067 ± 0.007 & 1.026 ± 0.009 & 1.076 ± 0.010 & 0.984 ± 0.005 & 2n11 & INT-G	\ \\
22244238+2230425 & 1.146 ± 0.021 & 1.067 ± 0.013 & 0.997 ± 0.016 & 1.065 ± 0.014 & 0.994 ± 0.013 & 0n01 & FLD-G	\ \\
SPECULOOS-2 & 1.056 ± 0.010 & 1.039 ± 0.007 & 1.009 ± 0.008 & 1.039 ± 0.009 & 0.985 ± 0.007 & 2n10 & INT-G	\ \\
SPECULOOS-3 & 1.139 ± 0.018 & 1.070 ± 0.011 & 1.011 ± 0.014 & 1.056 ± 0.013 & 0.993 ± 0.012 & 1n12 & INT-G	\ \\
LHS~3154 & 1.061 ± 0.016 & 1.056 ± 0.010 & 1.023 ± 0.013 & 1.054 ± 0.012 & 0.997 ± 0.010 & 1n02 & INT-G	\ \\
TRAPPIST-1(FIRE) & 1.105 ± 0.001 & 1.110 ± 0.009 & 1.084 ± 0.001 & 1.062 ± 0.001 & 0.951 ± 0.001 & 1n10 & INT-G	\ \\
TRAPPIST-1(SpeX) & 1.119 ± 0.001 & 1.093 ± 0.010 & 1.070 ± 0.002 & 1.070 ± 0.001 & 0.971 ± 0.001 & 1n01 & INT-G	\ \\
\enddata
\tablecomments{For TRAPPIST-1, we used the medium-resolution SpeX/SXD and FIRE spectra results derived by \cite{Gonzales2019}.}
\end{deluxetable*}

\section{Gravity Indicators: Methods and analysis}\label{Methods}

Our study focuses on gravity-sensitive features within spectral types M5.5–L0, as shown in \autoref{figure1} and \autoref{figure2}. Following the approach of \citetalias{Allers2013}, we calculated gravity-sensitive indices for the full sample, with results presented in \autoref{table2}. The methodology, based on \citetalias{Allers2013}, is detailed in \autoref{sec:gravity_method}.

We also measured the equivalent widths (EWs) of the K\,\textsc{i} lines at 1.1692, 1.1778, and 1.2529\,$\micron$ using the \texttt{equivalent width} function from the \texttt{Specutils} Python package \citep{Specutils2019}. Uncertainties on the EW measurements were estimated by scaling the mean flux uncertainty relative to the continuum level and propagating it across the line region, accounting for the number of spectral data points. Results are reported in \autoref{table_A1}.

\begin{figure*}[htbp!]
    \centering
    \includegraphics[width=\textwidth, height=0.4\textheight, keepaspectratio]{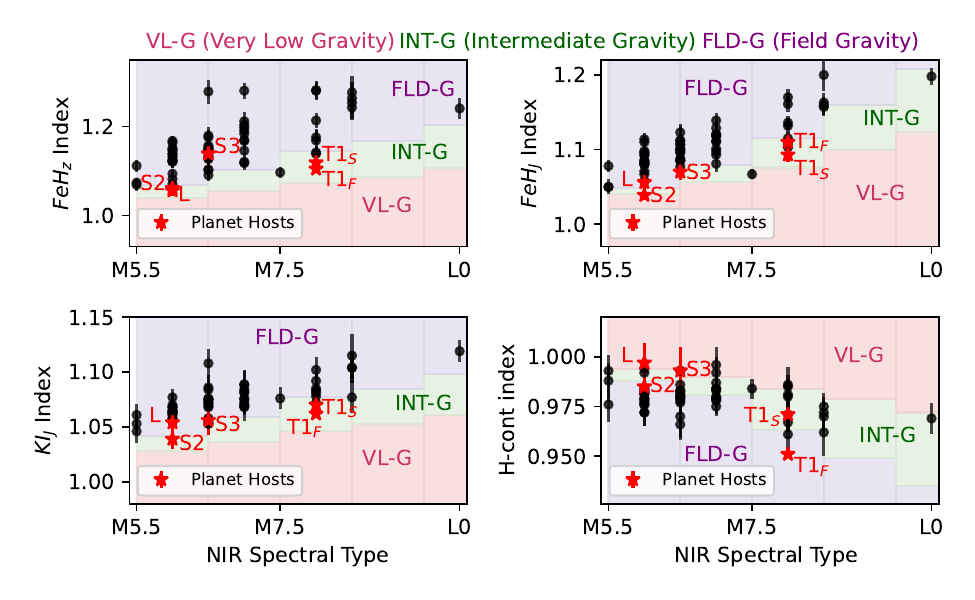} 
    \caption{Gravity indices (FeH$_z$, FeH$_J$, K\,\textsc{i}$_J$ and $H$-cont) vs.\ NIR spectral type for M dwarfs in our sample. The color-coded regions represent different gravity classifications: Very Low Gravity (VL-G, pink), Intermediate Gravity (INT-G, green), and Field Gravity (FLD-G, purple). SPECULOOS-2 (S2), SPECULOOS-3 (S3), LHS 3154 (L) and the FIRE and SpeX spectra measurements of TRAPPIST-1 ($\mathrm{T1}_{F}$ and $\mathrm{T1}_{S}$) are highlighted as red stars.}
    \label{figure3}
\end{figure*}

\begin{figure*}[htbp!]
    \centering
    \includegraphics[width=\textwidth, height=0.4\textheight, keepaspectratio]{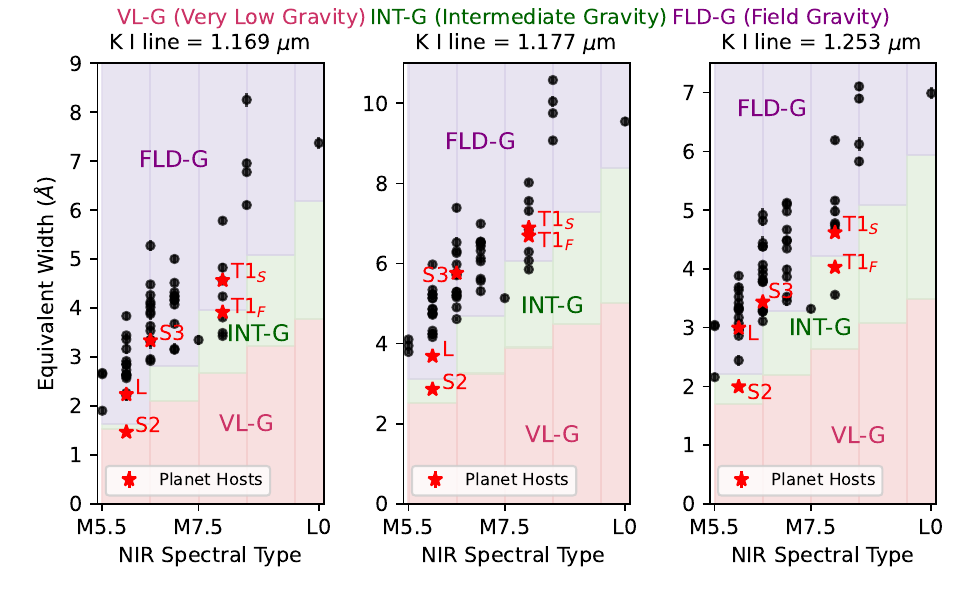}
    \caption{EWs for the K\,\textsc{i} lines in the J band vs. NIR spectral type for M dwarfs in our sample. The figure elements are the same as in \autoref{figure3}. The panels display: (a) NIR spectral type vs.\ K\,\textsc{i} 1.169\,$\micron$ EW, (b) NIR spectral type vs.\ K\,\textsc{i} 1.177\,$\micron$ EW, and (c) NIR spectral type vs.\ K\,\textsc{i} 1.253\,$\micron$ EW.}
     \label{figure4}
\end{figure*}

Based on the index values in \autoref{table2}, we assigned gravity classifications using the VL-G (very low gravity), INT-G (intermediate gravity), and FLD-G (field gravity) categories defined in \citetalias{Allers2013}, as illustrated in \autoref{figure3} and \autoref{figure4}.
Scores were assigned per index as follows: for FeH$_{J}$, FeH$_{Z}$, and KI$_{J}$, values below the VL-G threshold received a score of 2, between VL-G and INT-G a score of 1, and above INT-G a score of 0. For VO$_{z}$ and $H$-cont, scoring was reversed. A combined FeH score was derived by evaluating both FeH$_{Z}$ and FeH$_{J}$, taking the higher of the two. 

Whereas \citetalias{Allers2013} used a “?” designation when an index score was intermediate but had a $1\sigma$ uncertainty overlapping with field gravity, we followed the approach of  \cite{Aller2016} and \cite{Martin2017}, opting not to use this label. Instead, we used a Monte Carlo approach to more robustly quantify uncertainty in the gravity classification. For each index, we generated 10,000 synthetic values drawn from a Gaussian distribution defined by the measured mean and variance. From these realizations, we calculated a distribution of scores, and we report the mean and standard deviation to reflect both the central tendency and associated uncertainty.

The final gravity class was based on the median of the four main scores (FeH, VO\textsubscript{z}, KI\textsubscript{J}, H-cont): values $\leq 0.5$ were classified as FLD-G, between 0.5–1.5 as INT-G, and $\geq 1.5$ as VL-G.  \autoref{table2} presents the gravity scores and corresponding classifications for objects with spectral types. As in \cite{Gonzales2019}, half spectral types were rounded to the nearest integer.

\section{Results and Discussion}\label{results}

\subsection{Gravity Anomalies in Exoplanet-hosting M-Dwarfs}

\citet{Gonzales2019} compared TRAPPIST-1’s \citetalias{Allers2013} gravity indices across multiple datasets, including SpeX and FIRE spectra, and found a mix of youth-like and field-age features. More recently, \citet{Davoudi2024} applied a similar methodology using JWST/NIRISS data and confirmed intermediate-gravity (INT-G) classifications across all indices. Specifically, TRAPPIST-1 falls within the INT-G regime gravity indices are plotted against spectral type, but aligns more so with field gravity (FLD-G) in equivalent width comparisons. Certain indices, such as $H$-cont and KI$_{J}$, show field-like values in FIRE medium-resolution spectra. Together, these subtle but consistent deviations across datasets support an INT-G classification for TRAPPIST-1.
This interesting combination of field-age characterizations and low-gravity spectral features has been interpreted as evidence for external influences such as magnetic activity or tidal interactions with orbiting planets \citep{Gonzales2019}.

A similar pattern is observed in Teegarden’s Star, an M7.5 dwarf classified as a $\beta$ (intermediate-gravity) source by \citet{Gagne2015} based on SpeX/Prism spectra. \R{Its gravity indices---including a low FeH$_Z$ (1.07 $\pm$ 0.01), low KI$_J$ (1.05 $\pm$ 0.01), and elevated $H$-cont (1.00 $\pm$ 0.01)---deviate from typical field M7.5 dwarfs, further reinforcing the intermediate-gravity classification \citep{Gagne2015}.}

Our findings extend this trend to SPECULOOS-2, SPECULOOS-3, and LHS~3154.
Each shows spectral signatures indicative of intermediate gravity, supporting the hypothesis that such anomalies may be linked to the presence of close-in planetary companions. SPECULOOS-2 hosts two planets with orbital periods of 2.73 days (SPECULOOS-2~b) and 8.46 days (SPECULOOS-2~c) \citep{Delrez2022}. SPECULOOS-3 hosts a single planet ultra-short-period planet (0.72\,d; \citealt{Gillon2024}), and LHS~3154 hosts a planet with an orbital period of 3.72 days \citep{lhs3154}. Relevant spectral indices for these systems are presented in \autoref{figure3} and \autoref{figure4}.

\subsection{Assessing the Relationship between Surface Gravity and Planet Occurrence}

In this section, we evaluate whether surface gravity---quantified via overall classification or specific spectral features---shows a statistical relationship with the presence of close-in exoplanets in our sample.

\subsection{Gravity Classification and Planet Occurrence}

We investigated the relationship between spectroscopic gravity classification and exoplanet-hosting status among our sample of late-M dwarfs.
As an initial test, we applied Fisher’s exact test \citep{Fisher1922} under the null hypothesis that gravity class is independent of planet-hosting status.
The test returned a p-value of 0.0116, corresponding to a $2.27\sigma$ result, suggesting a marginally significant association.

However, our sample is not volume-complete and exhibits biases in the target selection, particularly for TRAPPIST-1, SPECULOOS-2, and LHS 3154, for which spectra were obtained only after planetary signals had been identified.
To mitigate the effects of this selection bias and sample incompleteness, we performed a regularized logistic regression using gravity class as a predictor for planet-hosting status, weighted by the inverse cube of Gaia distances \citep{GaiaCollaboration2023} to approximate a volume-limited correction.
We employed ridge (L2) regularization \citep{Hoerl1970, Pedregosa2011} to stabilize the fit in the presence of small-sample effects and potential quasi-separation.

The logistic regression results revealed no statistically significant increase in the likelihood of hosting a close-in planet for intermediate-gravity (INT-G) stars compared to field-gravity (FLD-G) stars.
This finding suggests that the apparent signal in the unweighted Fisher’s test may be influenced by the selection effects in the sample.
Thus, while the marginally significant result from the Fisher’s exact test is tantalizing, a more definitive assessment of the connection between gravity classification and exoplanet occurrence will require a volume-limited, unbiased spectroscopic survey of ultracool dwarfs.

In the following, we explore further whether any of the spectral indices related to the gravity classification show a unique relationship to known planet occurrence in our sample.

\subsection{Spectral Index Behavior and Planet Occurrence}


To statistically assess whether gravity-sensitive features differ between stars with and without \textit{detected} planets, we performed Kolmogorov--Smirnov (KS) tests \citep{Kolmogorov1933, Smirnov1948} on each gravity index from \autoref{table2}.
The results are summarized in \autoref{gravity_metallicity}.
We caution, however, that the sample includes a known selection bias: spectra for three of the known exoplanet host stars were obtained \textit{after} their exoplanets had been detected.
This targeted follow-up may skew the apparent differences between planet-hosting and non-hosting stars, complicating interpretation of any planet occurrence trends.

\begin{figure*}[htbp!]
    \centering
    \includegraphics[width=\textwidth]{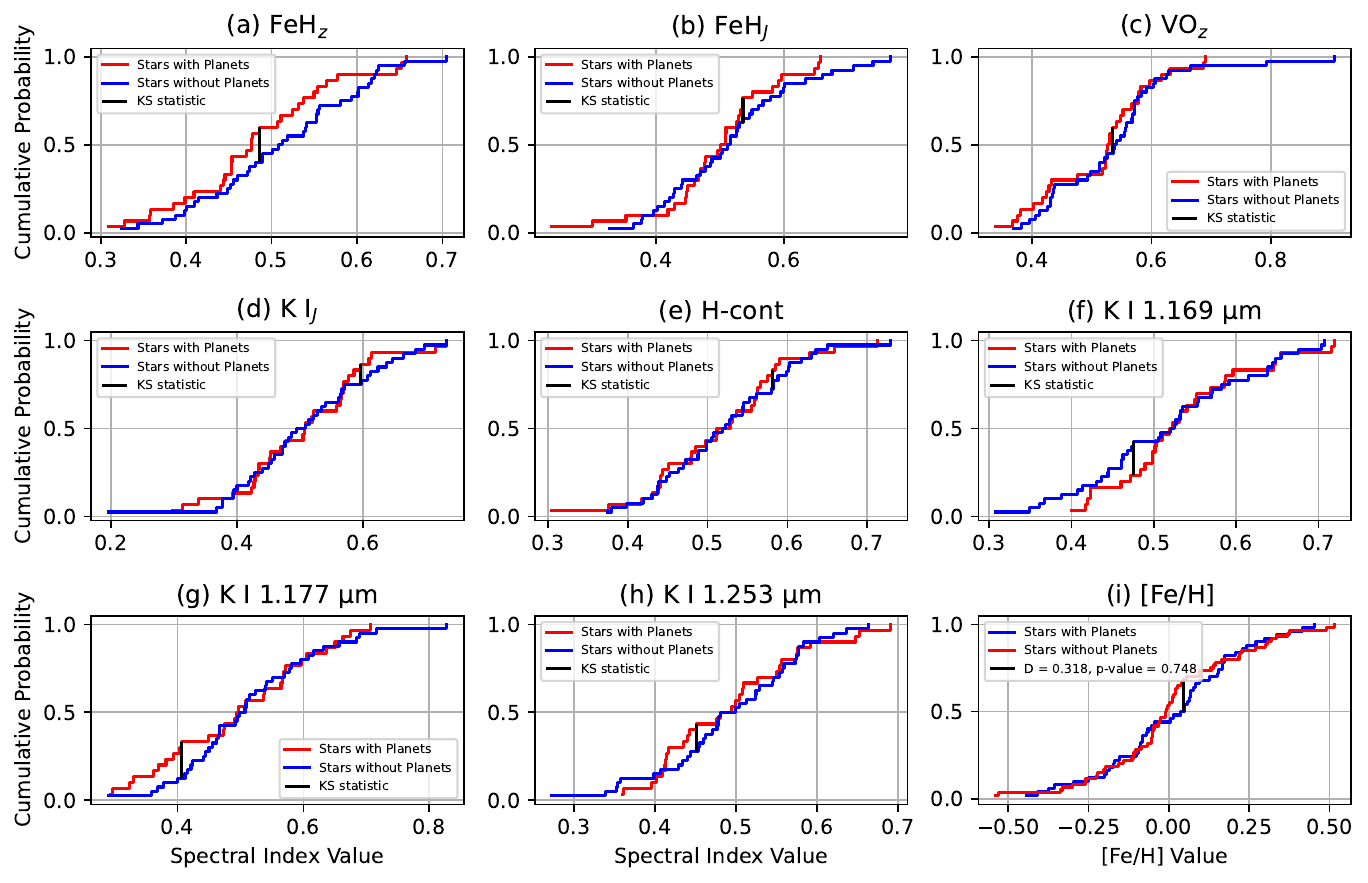}
    \caption{ECDF plots for eight gravity indices, comparing stars with and without detected planets using the Kolmogorov--Smirnov (KS) test. Each panel shows the empirical cumulative distribution functions (ECDFs) for the two samples, with the KS statistic ($D$) indicated by a solid vertical line representing the maximum vertical difference between the distributions. Panel (a) shows FeH$_z$, where a significant difference is observed between the two groups, as indicated by the larger KS statistic. For all other indices, including FeH$_J$, VO$_z$, K\,\textsc{i}$_J$, and metallicity, the distributions do not show significant differences ($p > 0.05$). The x-axis represents the spectral index values, and the y-axis represents the cumulative probability.}
     \label{figure5}
\end{figure*}

\autoref{figure5} presents empirical cumulative distribution function (ECDF) comparisons for eight gravity indices, comparing stars with and without detected transiting planets. However, we caution that stars classified as ``without planets'' may still host undetected planets, such as non-transiting systems or planets below the detection threshold. Therefore, the distinction between planet-hosting and non-hosting stars is not absolute, and any observed differences should be interpreted with care.

In each panel, the KS statistic ($D$) represents the maximum vertical distance between the two ECDFs, providing a quantitative measure of the discrepancy between the distributions. Among the indices tested, only FeH$_z$ exhibited a marginal difference between the two populations (KS statistic $D = 0.604$, $p = 0.043$), reaching the $2\sigma$ significance threshold. This may suggest a potential shift in its distribution associated with planet presence, though the result remained statistically tentative. All other indices, including FeH$_J$, VO$_z$, K\,\textsc{i}$_J$, and iron abundance [Fe/H], showed no statistically significant differences ($p > 0.05$).

To further investigate whether stars hosting detected transiting exoplanets differ in their gravity-sensitive features, we performed a chi-squared test on each gravity index.
The tests compared values from two groups, stars with detected exoplanets and stars without.
For FeH$_z$, the test yielded $\chi^2 = 9.342$ with a $p$-value of $0.052$, suggesting a marginally significant association with planet presence, though it does not reach the standard threshold for significance ($p < 0.05$). 
No such trend was observed for the other indices, with $\chi^2$ values ranging from $34.7$ to $65.0$ and corresponding $p$-values between 0.118 and 0.604.

\begin{table*}[ht!]
\caption{Statistical correlations between gravity indices and metallicity, with test statistic, $p$‑value, and Gaussian equivalent $\sigma$. Results with $\sigma > 3$ are bolded.}
\centering
\small
\begin{tabular}{lccccccccc}
\hline
Gravity Index & \multicolumn{3}{c}{KS test with planet presence} & \multicolumn{3}{c}{Pearson Correlation with [Fe/H]} & \multicolumn{3}{c}{Spearman Correlation with [Fe/H]} \\
 & $D$ & $p$ & $\sigma$ & $r$ & $p$ & $\sigma$ & $r_s$ & $p$ & $\sigma$ \\
\hline
FeH$_z$ & 0.604 & 0.043 & 2.02 & $-0.277$ & 0.031 & 2.15 & \textbf{$-0.458$} & \textbf{0.001} & \textbf{3.29} \\
FeH$_J$ & 0.511 & 0.129 & 1.52 & $-0.138$ & 0.289 & 1.06 & $-0.229$ & 0.075 & 1.78 \\
VO$_z$  & 0.364 & 0.471 & 0.72 & 0.288 & 0.025 & 2.24 & 0.348 & 0.006 & 2.74 \\
K\,\textsc{i}\,J & 0.550 & 0.083 & 1.73 & $-0.214$ & 0.097 & 1.66 & $-0.245$ & 0.057 & 1.90 \\
H-cont  & 0.386 & 0.402 & 0.83 & 0.160 & 0.217 & 1.23 & 0.187 & 0.149 & 1.44 \\
K\,\textsc{i}\,1.169\,\micron & 0.364 & 0.471 & 0.72 & $-0.133$ & 0.305 & 1.03 & $-0.231$ & 0.073 & 1.79 \\
K\,\textsc{i}\,1.177\,\micron & 0.400 & 0.353 & 0.93 & $-0.170$ & 0.191 & 1.30 & $-0.268$ & 0.051 & 1.96 \\
K\,\textsc{i}\,1.253\,\micron & 0.311 & 0.671 & 0.42 & $-0.195$ & 0.132 & 1.51 & $-0.322$ & 0.011 & 2.54 \\
\hline
\end{tabular}
\label{gravity_metallicity}
\end{table*}

\vspace{-0.7cm}
\subsection{Interplay Between Gravity and Metallicity Indicators in Late M-dwarf Spectra}

\subsubsection{Metallicity Estimation and Relationship to Planet Occurrence}

To explore potential trends between planet occurrence and metallicity ([Fe/H]), we calculated [Fe/H] values for all stars in our sample using the \texttt{splat} Python package\footnote{\url{https://gist.github.com/brackham/26b305919252f9bdabd2b56d463934d3}} \citep{Schneider2016, Splat}. For mid- and late-type M dwarfs, we estimated metallicities using the EWs of the Na\,\textsc{i} (2.205\,$\micron$) and Ca\,\textsc{i} (2.263\,$\micron$) absorption features, in combination with the H$_2$O--K2 index and the calibration provided by \citet{Mann2014}, following the method of \citet{Rojas2010, Rojas2012}. EWs were measured using the \texttt{Specutils} Python package, with uncertainties estimated via Monte Carlo simulations based on flux uncertainties \citep[see][]{Delrez2022}. A systematic uncertainty of 0.07\,dex from the \citet{Mann2014} calibration was added in quadrature to the measurement errors.

Our analysis revealed no significant correlation between [Fe/H] and planet occurrence within our sample, as indicated by the KS test (KS statistic = 0.318, p-value = 0.748; see panel~(i) of \autoref{figure5}). 

\subsubsection{Metallicity Dependence of Gravity-sensitive Indices}

Beyond planet occurrence, we also investigated whether gravity-sensitive spectral indices exhibit systematic trends with stellar metallicity.
To do so, we calculated both Pearson and Spearman correlation coefficients between [Fe/H] and a set of gravity-sensitive spectral indices (see \autoref{gravity_metallicity}). 

While Pearson’s test identifies strictly linear relationships, Spearman’s rank-based test is sensitive to broader monotonic trends, including non-linear or saturating behaviors.

Among the indices tested, FeH$_z$ and VO$_z$ showed the strongest correlations with metallicity.
However, only the Spearman test result for FeH$_z$ exceeded the $3\sigma$ threshold, indicating a robust inverse relationship between FeH$_z$ and [Fe/H].
This result reinforces the finding that FeH$_z$ is sensitive to metallicity in addition to gravity. 
VO$_z$ also showed moderate positive correlations with metallicity in both Pearson and Spearman tests, although only at the $2\sigma$ level. 
Similarly, the K\,\textsc{i} 1.253\,$\micron$ feature shows a weak anti-correlation with [Fe/H], though only at the $2\sigma$ level for the Spearman test.

\R{We note, however, that our analysis involved testing eight gravity-sensitive spectral indices for correlations with [Fe/H]. In such a context, the probability of finding at least one apparently significant $2\sigma$ (p < 0.05) result purely by chance increases due to multiple comparisons. For eight independent tests, this false-positive probability rises to approximately 34\% ($1 - 0.95^8$). The strongest result---an inverse correlation between FeH$_z$ and [Fe/H] at the 3.3$\sigma$ level---has a two-sided p-value of $\sim$0.001, corresponding to a $\sim$0.8\% chance of occurring at least once in eight independent tests ($1 - 0.999^8$). While this suggests the FeH$_z$ correlation is likely robust, the marginal trends seen in other indices (e.g., VO$_z$, K\textsc{i}) should be interpreted with caution unless confirmed by future studies or larger samples.}

Overall, the tendency for stronger correlations in the Spearman tests suggests that metallicity influences gravity-sensitive features in complex, non-linear ways, underscoring the need to consider both composition and gravity when interpreting M-dwarf spectra.

\subsubsection{Interpreting the FeH$_z$--[Fe/H] Anti-Correlation}

The strongest correlation we observe is an inverse relationship between FeH$_z$ and [Fe/H], which merits further interpretation given its potential implications for both spectral diagnostics and stellar characterization.
At face value, the negative correlation between FeH$_z$ and [Fe/H] is somewhat unexpected, as the FeH$_z$ index traces the strength of the FeH molecular feature, which is generally assumed to increase with iron abundance. 
However, the FeH$_z$ index is sensitive to both surface gravity and metallicity: lower gravity weakens FeH absorption, potentially offsetting the impact of metallicity. 
This dual sensitivity offers a plausible explanation for the observed anti-correlation, particularly if metal-rich stars tend to have low gravities due to either enhanced opacities throughout the photosphere or relative youth.

To test whether metallicity alone could explain differences in gravity classifications, we performed a Kruskal--Wallis H test \citep{Kruskal1952} comparing [Fe/H] across gravity classes in our sample.
The result was not statistically significant ($H = 1.99$, $p = 0.16$), indicating no strong evidence that [Fe/H] differs between FLD-G and INT-G objects in our sample.
This suggests that the relationship between surface gravity and metallicity in our sample is not straightforward, and thus the observed FeH$_z$--[Fe/H] anti-correlation may reflect the influence of other factors such as chemical abundances.
This reasoning is in line with the findings of \citet{Gonzales2019}, who found no evidence for low metallicity mimicking low-gravity spectral features for TRAPPIST-1. 

Another possible explanation for the FeH$_z$–[Fe/H] anti-correlation is that stellar activity may also modulate FeH absorption. In such cases, FeH$_z$ may not be a pure gravity indicator, but could reflect a combination of metallicity and activity-driven effects. For example, while low metallicity might intrinsically weaken FeH absorption, elevated activity levels, by contrast, could enhance it. This interplay could produce an inverse correlation between FeH$_z$ and [Fe/H], complicating efforts to disentangle gravity, metallicity, and activity influences on the spectral features.

\section{Conclusions}\label{conclusion}

\R{We present a comprehensive analysis of gravity-sensitive spectral indices in 57 ultracool M dwarfs, spanning spectral types M5.5 to L0 and including four exoplanet-hosting stars: TRAPPIST-1, SPECULOOS-2, SPECULOOS-3, and LHS~3154.
For the latter three targets, our analysis provides the first gravity classifications.}
Our study investigates the relationship between spectral gravity indices and exoplanet presence in late-M dwarfs. 
The main findings are as follows.

\begin{enumerate}
    \item We find that exoplanet-hosting M dwarfs, such as SPECULOOS-2, SPECULOOS-3, and LHS~3154, exhibit unusual spectral features similar to those of TRAPPIST-1 and Teegarden's Star. 
    These peculiarities suggest an intermediate-gravity classification, which is unexpected for field-age stars. 
    This finding raises the possibility that factors beyond stellar age may influence the gravity-sensitive spectral features observed in exoplanet-hosting M dwarfs. 
    
    One possible explanation we explore is that increased metallicity could inflate stellar radii by enhancing atmospheric opacity, thereby mimicking low-gravity signatures in spectral indices. 
    At the same time, higher metallicity could reasonably influence planet formation efficiency. 
    However, the connection between anomalous gravity indices and planet occurrence is not yet well constrained by current observations. 
    Nonetheless, this possible relationship underscores the need for caution when using spectral diagnostics to infer age and gravity in exoplanet-hosting ultracool dwarfs.

    \item We examined whether the presence of exoplanets correlates with specific gravity-sensitive spectral features, as tidal interactions or magnetic activity may modify these indices in exoplanet-hosting M dwarfs.
    Both KS and $\chi^2$ tests identified FeH$_z$ as the most promising index linked to planet presence.
    The KS test yielded $D = 0.604$ with $p = 0.043$, while the $\chi^2$ test returned $\chi^2 = 9.342$ and $p = 0.052$---both suggesting a possible trend near the $2\sigma$ level.
    The close agreement between these $p$-values reinforces the potential relevance of FeH$_z$, though these marginal results are insufficient to support a definitive correlation.
    Other gravity indices showed no significant differences between stars with and without detected planets.

    \item We also tested for a broader association between gravity class and planet occurrence.
    A Fisher's exact test yielded a marginal $2.3\sigma$ association between intermediate-gravity classification and planet-hosting status.
    However, this signal was not recovered in a volume-weighted logistic regression, indicating that it may be driven by selection biases in the sample.

    \item We observed strong correlations between metallicity and certain gravity-sensitive indices, particularly FeH$_z$ and VO$_z$.
    FeH$_z$ shows a robust inverse relationship with [Fe/H] (Spearman $r_s = -0.458$, $p = 0.001$, $3.3\sigma$), suggesting it is not solely a gravity indicator but is also sensitive to metallicity and potentially stellar activity.
    A Kruskal--Wallis H test revealed no significant difference in [Fe/H] between FLD-G and INT-G objects ($H = 1.99$, $p = 0.16$), implying that the FeH$_z$--[Fe/H] anti-correlation is not simply due to systematic metallicity differences across gravity classes. 
    This underscores the complex interplay between gravity, composition, and activity in shaping M-dwarf spectral features. 

\end{enumerate}

Given these findings, further research is needed to disentangle the effects of stellar activity, tidal interactions, and metallicity on the spectra of exoplanet-hosting stars. Stars with very short-period planets, such as SPECULOOS-3, may provide valuable insights into how close planetary orbits influence stellar spectra, especially in the context of magnetic interactions and activity levels.

\section*{Acknowledgments}
F. D. and M. G. acknowledge funding from the Belgian Federal Science Policy Office (BELSPO) for the BRAIN 2.0 project PORTAL (B2/212/P1/PORTAL).  They thank the European Space Agency (ESA) and BELSPO for their support in the framework of the PRODEX Programme. M. G. is FNRS-F.R.S. Research Director. 
B.V.R. thanks the Heising-Simons Foundation for support. 
J.d.W. and B.V.R. acknowledge support from the European Research Council (ERC) Synergy Grant under the European Union’s Horizon 2020 research and innovation program (grant No. 101118581 — project REVEAL).
This material is based upon work supported by the National Aeronautics and Space Administration under Agreement No.\ 80NSSC21K0593 for the program ``Alien Earths''. The results reported herein benefited from collaborations and/or information exchange within NASA’s Nexus for Exoplanet System Science (NExSS) research coordination network sponsored by NASA’s Science Mission Directorate.
Visiting Astronomer at the Infrared Telescope Facility, which is operated by the University of Hawaii under contract 80HQTR24DA010 with the National Aeronautics and Space Administration.
This paper includes data gathered with the 6.5 meter Magellan Telescopes located at Las Campanas Observatory, Chile.

%

\vspace{5mm}


\section*{Data Availability}

The data underlying Figures 1 and 2 are publicly available on Zenodo: \href{https://doi.org/10.5281/zenodo.16420633}{https://doi.org/10.5281/zenodo.16420633}. The dataset includes continuum-normalized near-infrared spectra in CSV format, containing columns for wavelength, flux, and flux uncertainty. A README file is also provided to describe the data structure. These materials are shared in accordance with AAS Journals' data-sharing policy.

\software{ \texttt{SPLAT} \citep{splat2, splat1}, Specutils \citep{Specutils2019} \texttt{NumPy} \citep{harris2020}, \texttt{Matplotlib} \citep{hunter2007}, \texttt{Astropy} \citep{AstropyCollaboration2022}, \texttt{emcee}\citep{Foreman2013}, \texttt{SciPy} \citep{Virtanen2020}}



\appendix
\restartappendixnumbering
\renewcommand{\thetable}{A\arabic{table}}

\section{Equivalent Widths and Metallicity Measurements}

\autoref{table_A1} presents the EW of the K\,\textsc{i} lines at 1.169\,$\micron$, 1.177\,$\micron$, and 1.253\,$\micron$, along with the metallicity ([Fe/H]) for each object in our sample.

\begin{deluxetable*}{llcccc}
\tablecaption{Equivalent widths and metallicity calculated from our sample spectra.\label{table_A1}}
\tablewidth{0pt}
\tabletypesize{\scriptsize}
\tablehead{
Object & Spectrum & EW(K\,\textsc{i} 1.169\,$\micron$)[\AA] & EW(K\,\textsc{i} 1.177\,$\micron$)[\AA]& EW(K\,\textsc{i} 1.253\,$\micron$)[\AA] & [Fe/H]
}
\startdata
2MASS J00202922+3305081 & SpeX & 1.902 ± 0.069 & 3.949 ± 0.066 & 2.156 ± 0.065 & 0.027 ± 0.114 \\
2MASS J00251602+5422547 & SpeX & 4.265 ± 0.062 & 6.263 ± 0.061 & 4.380 ± 0.055 & -0.188 ± 0.096 \\
2MASS J02195603+5919273 & SpeX & 2.673 ± 0.047 & 3.791 ± 0.045 & 3.045 ± 0.043 & 0.100 ± 0.084 \\
2MASS J02224767-2732349 & FIRE & 5.273 ± 0.112 & 7.390 ± 0.092 & 4.819 ± 0.084 & -0.156 ± 0.077 \\
2MASS J03544620+2416246 & SpeX & 2.211 ± 0.101 & 4.232 ± 0.101 & 2.442 ± 0.088 & 0.365 ± 0.103 \\
2MASS J04164276+1310587 & SpeX & 2.263 ± 0.052 & 4.852 ± 0.049 & 3.343 ± 0.046 & 0.219 ± 0.087 \\
2MASS J04333002+5635320 & SpeX & 3.490 ± 0.063 & 6.292 ± 0.061 & 4.740 ± 0.057 & -0.139 ± 0.102 \\
2MASS J04393407-3235516 & FIRE & 2.610 ± 0.067 & 4.164 ± 0.056 & 2.963 ± 0.052 & -0.005 ± 0.073 \\
2MASS J04490464+5138412 & SpeX & 3.916 ± 0.055 & 5.732 ± 0.051 & 3.775 ± 0.052 & 0.355 ± 0.093 \\
2MASS J04511406+0305285 & SpeX & 4.561 ± 0.100 & 6.846 ± 0.096 & 4.979 ± 0.087 & -0.109 ± 0.121 \\
2MASS J04511406+0305285 & FIRE & 5.002 ± 0.084 & 6.992 ± 0.069 & 5.124 ± 0.065 & 0.084 ± 0.075 \\
2MASS J04513734-5818519 & FIRE & 3.836 ± 0.073 & 5.976 ± 0.060 & 3.882 ± 0.057 & -0.006 ± 0.073 \\
2MASS J05220976+5754046 & SpeX & 6.103 ± 0.082 & 9.071 ± 0.077 & 5.829 ± 0.072 & 0.020 ± 0.100 \\
2MASS J05335379+5054170 & SpeX & 5.783 ± 0.069 & 8.021 ± 0.067 & 6.190 ± 0.063 & -0.230 ± 0.107 \\
2MASS J05512511+5511208 & SpeX & 2.916 ± 0.057 & 5.131 ± 0.053 & 3.420 ± 0.052 & 0.213 ± 0.092 \\
2MASS J06020172-1001565 & FIRE & 2.956 ± 0.077 & 4.906 ± 0.064 & 3.108 ± 0.061 & -0.005 ± 0.074 \\
2MASS J06431389+1631428 & SpeX & 4.071 ± 0.087 & 6.547 ± 0.083 & 4.488 ± 0.076 & -0.107 ± 0.116 \\
2MASS J06431389+1631428 & FIRE & 4.323 ± 0.068 & 6.437 ± 0.056 & 4.667 ± 0.053 & -0.146 ± 0.074 \\
2MASS J07552745-2404374 & FIRE & 3.632 ± 0.084 & 5.296 ± 0.070 & 3.785 ± 0.066 & 0.271 ± 0.074 \\
2MASS J08055713+0417035 & SpeX & 2.644 ± 0.049 & 4.746 ± 0.047 & 2.861 ± 0.045 & 0.145 ± 0.087 \\
2MASS J08055713+0417035 & FIRE & 2.918 ± 0.057 & 4.612 ± 0.048 & 3.309 ± 0.045 & 0.100 ± 0.072 \\
2MASS J08330310+3706083 & SpeX & 3.429 ± 0.060 & 5.852 ± 0.057 & 3.558 ± 0.056 & 0.441 ± 0.096 \\
2MASS J08334323-5336417 & FIRE & 3.677 ± 0.073 & 5.567 ± 0.061 & 3.865 ± 0.057 & 0.045 ± 0.073 \\
2MASS J09332510-4353384 & FIRE & 3.165 ± 0.102 & 5.231 ± 0.085 & 3.640 ± 0.081 & -0.093 ± 0.077 \\
2MASS J09332625-4353366 & FIRE & 3.360 ± 0.102 & 5.343 ± 0.084 & 3.686 ± 0.081 & -0.083 ± 0.076 \\
2MASS J09365564-2609422 & SpeX & 4.236 ± 0.071 & 7.560 ± 0.069 & 5.161 ± 0.063 & -0.203 ± 0.094 \\
2MASS J09432994-3833560 & SpeX & 3.441 ± 0.056 & 5.142 ± 0.053 & 3.320 ± 0.051 & -0.300 ± 0.093 \\
2MASS J10424135-2416050 & SpeX & 3.311 ± 0.141 & 5.282 ± 0.131 & 3.909 ± 0.132 & -0.180 ± 0.277 \\
2MASS J10542786-5431322 & FIRE & 3.877 ± 0.123 & 5.706 ± 0.102 & 3.819 ± 0.099 & 0.343 ± 0.079 \\
2MASS J11155037-6731332 & FIRE & 4.825 ± 0.097 & 7.314 ± 0.080 & 4.664 ± 0.075 & 0.396 ± 0.075 \\
2MASS J11231964-0509045 & SpeX & 2.622 ± 0.063 & 4.720 ± 0.061 & 3.171 ± 0.058 & 0.044 ± 0.102 \\
2MASS J12294530+0752379 & SpeX & 4.014 ± 0.054 & 6.307 ± 0.052 & 3.985 ± 0.049 & -0.089 ± 0.089 \\
2MASS J13273095+0149384 & SpeX & 3.535 ± 0.064 & 5.977 ± 0.062 & 4.069 ± 0.057 & -0.071 ± 0.091 \\
2MASS J13313937-6513056 & FIRE & 7.371 ± 0.132 & 9.545 ± 0.112 & 6.990 ± 0.104 & 0.029 ± 0.079 \\
2MASS J14230252+5146303 & SpeX & 4.252 ± 0.078 & 6.056 ± 0.075 & 5.097 ± 0.070 & -0.218 ± 0.130 \\
2MASS J14253465+2540050 & SpeX & 2.567 ± 0.053 & 4.322 ± 0.051 & 2.954 ± 0.050 & 0.099 ± 0.090 \\
2MASS J16105843-0631325 & SpeX & 2.648 ± 0.053 & 4.104 ± 0.052 & 3.025 ± 0.050 & -0.038 ± 0.088 \\
2MASS J16210447-3711373 & SpeX & 4.515 ± 0.074 & 6.130 ± 0.071 & 4.477 ± 0.064 & 0.028 ± 0.095 \\
2MASS J17120433-0323300 & SpeX & 2.641 ± 0.051 & 4.758 ± 0.050 & 3.301 ± 0.046 & 0.161 ± 0.085 \\
2MASS J17364180-3425459 & FIRE & 2.848 ± 0.095 & 5.225 ± 0.079 & 3.073 ± 0.074 & -0.938 ± 0.076 \\
2MASS J17415439+0940537 & SpeX & 4.154 ± 0.062 & 6.321 ± 0.060 & 4.978 ± 0.060 & -0.120 ± 0.102 \\
2MASS J18365842-3507176 & FIRE & 4.078 ± 0.152 & 6.527 ± 0.126 & 4.920 ± 0.116 & -0.403 ± 0.080 \\
2MASS J18485108-8214422 & FIRE & 3.165 ± 0.111 & 5.310 ± 0.092 & 3.454 ± 0.088 & 0.103 ± 0.077 \\
2MASS J18545092-5704417 & FIRE & 4.170 ± 0.103 & 6.509 ± 0.085 & 3.989 ± 0.082 & 0.198 ± 0.076 \\
2MASS J19212977-2915507 & SpeX & 4.230 ± 0.049 & 6.534 ± 0.048 & 4.346 ± 0.046 & 0.177 ± 0.087 \\
2MASS J19332754+2150009 & SpeX & 6.776 ± 0.086 & 9.753 ± 0.082 & 7.106 ± 0.075 & -0.082 ± 0.104 \\
2MASS J19395199-5750339 & FIRE & 4.120 ± 0.169 & 5.766 ± 0.140 & 4.439 ± 0.126 & 0.182 ± 0.081 \\
2MASS J19544358+1801581 & SpeX & 3.808 ± 0.061 & 6.072 ± 0.059 & 4.776 ± 0.055 & 0.530 ± 0.088 \\
2MASS J20125255+1246315 & SpeX & 4.482 ± 0.080 & 5.156 ± 0.077 & 4.380 ± 0.067 & 0.046 ± 0.104 \\
2MASS J20291194+5750317 & SpeX & 2.849 ± 0.065 & 4.247 ± 0.065 & 2.997 ± 0.061 & 0.157 ± 0.101 \\
2MASS J20495272-1716083 & SpeX & 3.344 ± 0.097 & 5.198 ± 0.092 & 3.281 ± 0.088 & 0.046 ± 0.129 \\
2MASS J21010483+0307047 & SpeX & 3.350 ± 0.097 & 5.210 ± 0.092 & 3.271 ± 0.088 & 0.047 ± 0.129 \\
2MASS J21265788+2531080 & SpeX & 6.955 ± 0.083 & 10.580 ± 0.078 & 6.898 ± 0.072 & -0.212 ± 0.107 \\
2MASS J21381698+5257188 & SpeX & 3.147 ± 0.065 & 5.618 ± 0.062 & 3.525 ± 0.059 & -0.050 ± 0.106 \\
2MASS J21513137-4017229 & FIRE & 3.348 ± 0.094 & 5.135 ± 0.078 & 3.319 ± 0.074 & 0.236 ± 0.076 \\
2MASS J22244238+2230425 & SpeX & 2.734 ± 0.092 & 4.871 ± 0.089 & 3.525 ± 0.083 & -0.116 ± 0.111 \\
SPECULOOS-2 & SpeX & 1.466 ± 0.052 & 2.860 ± 0.052 & 1.993 ± 0.051 & -0.028 ± 0.089 \\
SPECULOOS-3 & SpeX & 3.335 ± 0.080 & 5.760 ± 0.076 & 3.436 ± 0.072 & 0.070 ± 0.100 \\
LHS~3154 & SpeX & 2.233 ± 0.069 & 3.687 ± 0.069 & 2.993 ± 0.065 & 0.232 ± 0.097 \\
TRAPPIST-1 & FIRE & 3.913 ± 0.019 & 6.690 ± 0.014 & 4.027 ± 0.014 & 0.052 ± 0.073 \\
TRAPPIST-1 & SpeX & 4.566 ± 0.084 & 6.891 ± 0.073 & 4.618 ± 0.067 & 0.052 ± 0.073 \\
\enddata
\tablecomments{For TRAPPIST-1, we used the medium-resolution SpeX/SXD and FIRE spectra results derived by \cite{Gonzales2019} and its [Fe/H] determined by \cite{Davoudi2024}.}
\end{deluxetable*}

\section{Measuring Surface Gravity Indicators}\label{sec:gravity_method}

We adopt the spectral index method developed by \citetalias{Allers2013}, which uses a three-bandpass formula to quantify gravity-sensitive absorption features:

\begin{equation}
\text{index} = \left( 
\frac{\lambda_{\text{line}} - \lambda_{\text{cont1}}}{\lambda_{\text{cont2}} - \lambda_{\text{cont1}}} F_{\text{cont2}} +
\frac{\lambda_{\text{cont2}} - \lambda_{\text{line}}}{\lambda_{\text{cont2}} - \lambda_{\text{cont1}}} F_{\text{cont1}} 
\right) / F_{\text{line}}.
\end{equation}

Here, $\lambda$ refers to the central wavelengths of the continuum and line regions, and $F$ is the average flux in each band. Band definitions are adopted from Table~4 of \citetalias{Allers2013}, with one exception: for the H-cont index, we used a revised blue continuum window at $1.270\,\micron$ due to spectral coverage limits, ensuring consistency with the original index’s sensitivity.

Index values are converted into discrete gravity scores (0--2) using spectral-type-dependent thresholds from Table~9 of \citetalias{Allers2013}. For instance, an M7 object scores 2 for FeH\textsubscript{z} if $\mathrm{FeH}_z \leq 1.056$.



\bibliography{sample631}{}

\bibliographystyle{aasjournal}



\end{document}